\documentclass[fleqn,10pt]{wlscirep}
\usepackage[utf8]{inputenc}
\usepackage[T1]{fontenc}
\usepackage{amsmath, array, amssymb, physics}
\usepackage{xr}
\usepackage{url}
\newcommand{\euler}{\mathrm{e}}
\newcommand{\im}{\mathrm{i}}
\newcommand{\bessel}{\mathrm{J}}

\title{Propagation of optically tunable coherent radiation in a gas of polar molecules}

\author[1,*]{Piotr G\l{}adysz}
\author[]{Piotr Wcis\l{}o}
\author[]{Karolina S\l{}owik}
\affil[]{Institute of Physics, Faculty of Physics, Astronomy and Informatics, Nicolaus Copernicus University in Toru\'{n}, Grudziadzka 5, 87-100 Torun, Poland}

\affil[*]{glad@doktorant.umk.pl}

\begin{abstract}
Coherent, optically dressed media composed of two-level molecular systems without inversion symmetry are considered as all-optically tunable sources of coherent radiation in the microwave domain. A theoretical model and a numerical toolbox are developed to confirm the main finding: the generation of a low-frequency radiation, and the buildup and propagation dynamics of such low-frequency signals in a medium of polar molecules in a gas phase. The physical mechanism of the signal generation relies on the permanent dipole moment characterizing systems without inversion symmetry. The molecules are polarized with a DC electric field yielding a permanent electric dipole moment in the laboratory frame; the direction and magnitude of the moment depend on the molecular state. As the system is resonantly driven, the dipole moment oscillates at the Rabi frequency and, hence, generates microwave radiation. We demonstrate the tuning capability of the output signal frequency with the drive amplitude and detuning. We find that even though decoherence mechanisms such as spontaneous emission may damp the output field, a scenario based on pulsed illumination yields a coherent, pulsed output of tunable temporal width. Finally, we discuss experimental scenarios exploiting rotational levels of gaseous ensembles of heteronuclear diatomic molecules.

\end{abstract}

\begin{document}

\flushbottom
\maketitle

\thispagestyle{empty}

\section*{Introduction}

Analysis of the propagation of an electric field in various types of media is a well-known problem in quantum optics, especially in the context of coherent phenomena \cite{mandel1995}. Typically, an ensemble of atoms is dressed with a coherent electromagnetic beam modifying the atomic optical properties probed by a weak beam of light. Attractive examples include stimulated Raman adiabatic passage \cite{drummond2002,vitanov2017}, magneto-optical rotation \cite{Budker2000,petrosyan2004,Pustelny2006} electromagnetically induced transparency \cite{harris1997,paspalakis2000,fleischhauer2005}, light slowdown \cite{hau1999,fleischhauer2000dark} and storage \cite{phillips2001,raczynski2007,slowik2012}.
Coherent atomic systems support quantum interference in emission or absorption channels, exploited e.g. in the phenomenon of lasing without inversion\cite{scully1989,harris1989,kocharovskaya1992}. An alternative scenario for lasing closely related to this work is based on a resonant amplification of a high-frequency signal generated in a superradiant  ensemble of atoms coherently driven by a low-frequency beam\cite{svidzinsky2013,shchedrin2015}. Despite the long record of investigation, there is still room in the field for fundamental and uncharted research ideas for even the simplest possible models. One such example is a one-dimensional medium consisting of two-level systems with broken inversion symmetry, coupled to an electromagnetic field. 

A two-level system, obeying inversion symmetry, driven by a resonant electromagnetic field is a canonical example considered in quantum optics. Such a system undergoes Rabi oscillations of populations of the two levels as the system subsequently exchanges energy with the driving field. This induces oscillations of a transition dipole moment at the transition frequency. This can be understood in terms of AC Stark effect. The resulting emission spectrum has the form of a Mollow triplet\cite{grove1977}, with a central peak at the transition frequency of the atomic system and sidebands detuned by the Rabi frequency. If the two-level system is polar, its inversion symmetry is broken and radiation is additionally generated at the Rabi frequency, i.e., much below the transition frequency between the eigenstates. This can be explained by noting that the permanent electric dipole moment characterizing the system can, in general, be different in the excited and ground states (polar molecules have to be subjected to an external DC electric field to have the electric dipole moment in the laboratory frame). Thus, the population exchange between the eigenstates results in oscillations of the permanent dipole moment at the Rabi frequency. 
Such behaviour was first shown by Kibis et al.\cite{kibis2009}, where the coupling of a single asymmetric system (a quantum dot) with classical light was considered.
The discussion was extended to the cases of single-mode~\cite{savenko2012} and bichromatic light\cite{kryuchkyan2017}. Nonlinear effects were also considered in terms of polarizabilities of asymmetric systems~\cite{Paspalakis2013}. Moreover, two-level systems with broken inversion symmetry were suggested for lasing or for generation of squeezed light~\cite{marthaler2016,koppenhofer2016, chestnov2017}.

Until now, the analysis neglected field propagation effects. This approach limited the possible physical realizations to small amounts of relatively large quantum systems and excluded commonly available molecular clouds or solids. In this paper, we extend that research to scenarios in which a resonant driving beam illuminates an ensemble of two-level systems with broken inversion symmetry, e.g., polar molecules in a gaseous phase. As a result, a coherent beam of low-frequency radiation is generated throughout the length of the sample and propagates parallel to the drive.

We investigate the performance of the proposed system as an all-optically-controlled source of coherent radiation. Its frequency could be optically tuned within the microwave domain by modulation of the amplitude of the drive. In an opposite strategy, modulation of the driving field's detuning from resonance with the atomic transition would provide a knob to suppress the outgoing signal at a fixed frequency. To investigate these effects, we develop a method based on the semiclassical Bloch--Maxwell equations derived under a generalized form of the rotating wave approximation. The usual slowly varying envelope approximation is not applied because it does not hold for low-frequency pulses. The theory is applied to a realistic model of a gaseous medium of molecules, which support a permanent dipole moment.

\section*{Method}

\subsection*{Medium}\label{sec:medium}
We consider a quasi-one-dimensional sample containing a medium of uniformly distributed polar molecules. To generate a signal the dipole moments have to be oriented in the laboratory frame which can be done by applying an external DC electric field. The simplest examples of molecules well suited for this purpose include heteronuclear molecules with large electric dipole moment which can be relatively easily polarized in the laboratory frame, such as: methylidyne (CH)\cite{takezaki1995} or carbon fluoride (CF)\cite{weibel1997} both in a ground state $X\;^2\Pi_{1/2}$, hydroxide (OH)\cite{hain1997} in a $X\;^2\Pi_{3/2}$ state, or lithium hydride (LiH)\cite{dagdigian1979lih} in a $X\;^1\Sigma^+$ state. For such systems the dipole moment operator in the laboratory frame $\hat{\mathbf{d}}=\sum_{ij}\mathbf{d}_{ij}\ket{i}\bra{j}$ with $i,j\in\{e,g\}$ has nonzero diagonal elements: $\left|\mathbf{d}_{ii}\right|\neq 0$ (for a proof, see the Supplementary Information). The diagonal elements correspond to \emph{permanent} dipole moments, in contrast to the usually considered off-diagonal elements describing \emph{transition} dipole moments $\mathbf{d}_{ij}$ with $i\neq j$. A crucial for this work feature is that the permanent dipole moments of the two eigenstates are not equal: $\mathbf{d}_{ee}\neq\mathbf{d}_{gg}$. Naturally, the transition dipole moment $\mathbf{d}_{eg}=\mathbf{d}_{ge}^\star$ should be nonzero to enable efficient coupling with the driving electric field, which we describe in the following subsection. For the purpose of describing how the DC electric field polarizes the molecules in the laboratory frame we use the full set of relevant molecular levels, see Supplementary Information. However, for the purpose of describing the light-molecules interaction and light propagation it suffices to focus only on the two levels coupled by light (the drive beam), hence the medium is described with $2\times 2$ density matrix $\rho(z,t)$ dependent on position $z$ and time $t$, which we describe on the basis of excited ($e$) and ground ($g$) states $\{\ket{e}, \ket{g}\}$. The free Hamiltonian of the system can be written as $H_\mathrm{medium}=\hslash \omega_0 \ket{e}\bra{e}$, where $\hslash$ stands for the reduced Planck constant, $\omega_0$ is the transition frequency, and we have set the energy of the lower state to zero.

\subsection*{Electromagnetic field}\label{sec:field}
The electromagnetic field is treated classically. It consists of two components corresponding to the driving and signal fields (in this section we skip the DC electric field that is used to polarize the molecular medium). Thus, the electric part of the field can be written as
\begin{equation}
    \mathbf{E}(z,t)=\mathbf{E}_\mathrm{drive}(z,t)+\mathbf{E}_\mathrm{signal}(z,t).
\label{eq:ffield}    
\end{equation}
The drive
\begin{equation}
    \mathbf{E}_\mathrm{drive}(z,t)={\mathcal{E}}(z,t)\mathbf{e}\cos(kz-\omega t)
    \label{eq:dfield}    
\end{equation} 
is a strong coherent laser beam of carrier frequency $\omega$ close to resonance with the two-level system's transition frequency $\omega_0$.
Depending on the system, it may belong to the optical, near- or even far-infrared regime. Here, $\mathbf{e}$ indicates the polarization of the driving field, which we assume is constant. The beam propagates along the $z$ direction, and on resonance the wavenumber $k$ equals $\frac{\omega}{c}$ with $c$ being the speed of light in vacuum.
Tuned close to the resonance with the two-level systems, the driving field induces coherent Rabi oscillations of their population between the ground and excited states. Either due to back-action from the medium or due to external tuning, the envelope $\mathcal{E}$ can be modulated at timescales comparable to the inverse Rabi frequency. This observation will be important in later parts of this work. 

Due to coherent oscillations of the population between the eigenstates with unequal permanent dipole moments, a coherent signal field $\mathbf{E}_\mathrm{signal}(z,t)$ may be generated in the medium. Its source is the permanent dipole oscillating at the Rabi frequency $\Omega_R$, whose exact form is derived later. It is typically in the microwave regime.
Formally, we can distinguish the signal field envelope and a harmonic term
\begin{equation}
    \mathbf{E}_\mathrm{signal} = \mathcal{E}_\mathrm{s}(z,t)\mathbf{e}_\mathrm{s}e^{\im\Omega_R(\frac{z}{c}- t)}+\text{c.c.},
\end{equation}
where $\mathbf{e}_\mathrm{s}$ represents the polarization vector of the signal field, which is constant and parallel to the permanent dipole moments of the medium. 
In realistic scenarios, the timescales of modulations of the signal field envelope may be comparable to the inverse Rabi frequency $\left|\frac{\partial}{\partial t}\mathcal{E}_\mathrm{s}\right|\lesssim\mathcal{E}_\mathrm{s}\Omega_R$, which means that the slowly varying envelope approximation may not be valid for this part of the field and will not be applied. We make the reasonable assumption that initially the signal field is absent throughout the entire sample $\left|\mathbf{E}_\mathrm{signal}(z,t=0)\right|=0$ for all $z\in [0,L]$, where $L$ is the sample length. 
In the following section, we derive the equations describing the buildup and propagation of the signal field.

\subsection*{Medium coupled to the field}\label{sec:coupling}
The goal of this subsection is to find the coupled Bloch-Maxwell equations that govern the evolution of the density matrix of the medium as well as the propagation of the signal field. In many cases, the propagation effects for the drive can be neglected. Here, we will nevertheless allow temporal tuning of the drive amplitude at reasonable timescales: $\left|\frac{\partial}{\partial t}\mathcal{E}\right|\ll\mathcal{E}\omega$.

Our starting point is the one-dimensional form of the propagation equation for the total field given by Eq.~(\ref{eq:ffield})
\begin{equation}
    -\frac{\partial^2}{\partial z^2}\mathbf{E}(z,t)+\frac{1}{c^2} \frac{\partial^2}{\partial t^2} \mathbf{E}(z,t)=-\mu_0\frac{\partial^2}{\partial t^2}\mathbf{P}(z,t),
\label{eq:wave}
\end{equation}
where $\mu_0$ is the vacuum permeability, and $\mathbf{P}(z,t)$ is the polarization of the medium. For this equation to be valid, the medium must be nonmagnetic, linear, and isotropic, without free charges and currents. We make these assumptions here as they describe a rather wide class of molecular vapors and solids.

The polarization at the right-hand-side of Eq.~(\ref{eq:wave}) can be expressed as
\begin{equation}
    \mathbf{P}(z,t)= N\Tr(\rho(z,t)\mathbf{d})=N\sum_{i,j\in\{e,g\}}\rho_{ij}(z,t)\mathbf{d}_{ji},
\label{eq:polarization}
\end{equation}
where $N$ is the concentration of the two-level systems in the medium, $\Tr(\cdot)$ stands for the trace operation, and $\rho_{ij}(z,t)$ are elements of the density matrix of the medium $\rho(z,t)$. Note that contrary to the usually considered symmetric case, all density matrix elements contribute to the polarization, including the diagonal ones that give rise to the signal buildup. 

To find the explicit form of the source term in Eq.~(\ref{eq:wave}), we model the evolution of the density matrix with the master equation
\begin{equation}
    \im\hslash \frac{\partial}{\partial t}\rho(z,t)=[H(z,t) , \rho(z,t)] + \mathcal{L}\left[\rho(z,t)\right],
\label{eq:master}    
\end{equation}
where $H$ is the full Hamiltonian  
\begin{equation}
    H(z,t)=H_\mathrm{medium}-\mathbf{d}\cdot\mathbf{E}(z,t),
\label{eq:hamiltonian}    
\end{equation}
$H_\mathrm{medium}$ was introduced at the end of subsection \textit{Medium}, and the electric dipole approximation has been assumed for the interaction term. 
The relaxation term is given by \cite{lindblad1976,Gorini1976}
\begin{equation}
    \mathcal{L}\left[\rho(z,t)\right] = 2\im \hslash\sum_{p=\mathrm{se},\mathrm{coll}}\gamma_p\left(L_p\rho L_p^\dagger-\frac{1}{2}\rho L_p^\dagger L_p-\frac{1}{2} L_p^\dagger L_p\rho\right),
\end{equation} 
where in our system the index $p="\mathrm{se}"$ corresponds to spontaneous emission with $L_\mathrm{se}=\ket{g}\bra{e}$, while $p="\mathrm{coll}"$ describes collisional relaxation: $L_\mathrm{coll}=\ket{e}\bra{e}-\ket{g}\bra{g}$. To simplify the notation, we have omitted the arguments of the density matrix on the right-hand side.

From now on we assume that all matrix elements of the dipole moment operator $\mathbf{d}_{ij}$ are oriented in the same direction, parallel to the polarization direction of the fields. Hence, we can omit the vector notation. Please note, however, that the analysis could be equivalently performed for perpendicular orientations of permanent and transition dipoles: $\mathbf{d}_{ii} \perp \mathbf{d}_{ij}$, $j\neq i$. 

To separate slowly varying and rapidly-oscillating components of the coherence $\rho_{eg}$, we adjust the ansatz introduced in the previous work\cite{kibis2009}
\begin{equation}
    \rho_{eg(z,t)}=r_{eg}(z,t)\euler^{\im(kz-\omega t)}\euler^{-\im\kappa(z,t)\sin(kz-\omega t)},
\label{eq:ansatz}
\end{equation}
where the dimensionless parameter $\kappa(z,t)=\frac{\mathcal{E}(z,t)(d_{ee}-d_{gg})}{\hslash \omega}$ is a measure of asymmetry. As we will see from the Bloch equations, the timescale for variations of the envelope function $r_{eg}$ is on the order of $\Omega_R^{-1}$.

The last exponent in Eq.~(\ref{eq:ansatz}) can be expressed using the identity $\euler^{-\im\kappa\sin x}=\sum_{n=-\infty}^{+\infty}\bessel_n(\kappa)\euler^{-\im nx}$, where $\bessel_n$ refers to the Bessel function of the first kind. Making use of this form and inserting Eqs.~(\ref{eq:hamiltonian}--\ref{eq:ansatz}) into the master equation (\ref{eq:master}), with the field in the form given by Eq.~(\ref{eq:ffield}), we arrive at a set of Bloch equations whose general form and detailed derivation are given in the Supplementary Information. The rotating wave approximation leads to the following form of the Bloch equations:
\begin{subequations}
\begin{equation}
    \frac{\partial}{\partial t}\rho_{ee}
    =2\Im(\Omega^\star_Rr_{eg})+2\frac{E_\mathrm{signal}}{\hslash}\bessel_1(\kappa)\Im(d^\star_{eg}r_{eg})-2\gamma_\mathrm{se} \rho_{ee},
\label{eq:blocha}    
\end{equation}
\begin{equation}
    \frac{\partial}{\partial t}r_{eg}
    =\im\left[-\delta+\frac{\partial}{\partial t}\kappa+\frac{E_\mathrm{signal}}{\hslash} (d_{ee}-d_{gg})\right]r_{eg}+\im\left[\Omega_R+\frac{E_\mathrm{signal}}{\hslash}\bessel_1(\kappa)d_{eg}\right] (1-2\rho_{ee})-(\gamma_\mathrm{se}+\gamma_\mathrm{coll}) r_{eg},
\label{eq:blochb}
\end{equation}
\label{eq:bloch}
\end{subequations}
\noindent where we have introduced the detuning \mbox{$\delta=\omega_0 - \omega \ll \omega_0$} and derived the explicit form of the Rabi frequency \mbox{$\Omega_R=\frac{d_{eg} \omega}{d_{ee}-d_{gg}} \bessel_1(\kappa)$}. This form implies that the signal frequency can be controlled with external parameters, in particular the amplitude of the drive, as well as - to a smaller extent - its frequency. Note that if the diagonal and off-diagonal terms of the dipole moment are comparable, the order of magnitude for $\kappa$ corresponds to the ratio of the Rabi frequency to the transition frequency in the system. This means that $\kappa$ is typically small, and the Bessel function can be approximated by the linear term $\bessel_1(\kappa)\approx \frac{1}{2}\kappa$. In this regime, the Rabi frequency reverts to the familiar form $\Omega_R\approx \frac{\mathcal{E}(z,t)d_{eg} }{2\hslash}$. It is now clear that the elements $\rho_{ee}$, $r_{eg}$ of the density matrix vary at timescales set by $\Omega_R^{-1}$ and the inverse decoherence rates $\gamma^{-1}_p$. From Eq.~(\ref{eq:blochb}), it follows that the signal field gives rise to a time- and space-dependent energy shift of the levels that is proportional to the difference between the permanent dipole moments of the eigenstates. In addition, the time dependence of the envelope of the driving field is included in the term $\frac{\partial}{\partial t}\kappa$. This part is negligible for slow changes of the drive's envelope and vanishes for a constant amplitude. 

Once we have found the evolution of the density matrix elements, we can specify the form of the medium polarization induced by the driving field. Making use of the density matrix normalization $\rho_{gg}+\rho_{ee}=1$, combining Eqs.~(\ref{eq:ansatz}) and (\ref{eq:polarization}) yields 
\begin{equation}
    P(z,t) = N\Big[d_{gg}+\rho_{ee}(d_{ee}-d_{gg})+r_{eg}d_{ge}\sum_{n=-\infty}^\infty\bessel_n(\kappa)\euler^{-\im (n-1)(kz-\omega t)}+\text{c.c.} \Big].
\label{eq:polarization_full}    
\end{equation}
We can insert this form into the right-hand side of Eq.~(\ref{eq:wave}). To extract the equation for the signal rather than the total field, we insert Eq.~(\ref{eq:ffield}) in the left-hand-side of Eq.~(\ref{eq:wave}). Next, we separate contributions on both sides of the wave equation that oscillate at different harmonic frequencies. This step is based on the assumption that the components of the field oscillating at the Rabi frequency that contribue to the signal have their source in the polarization components oscillating at the Rabi frequency, while their coupling with the polarization components at the carrier frequency is negligible \cite{Paspalakis2013}. The details of this step are given in the Supplementary Information. As a result, we obtain the propagation equation of the signal field:
\begin{equation}
\begin{split}
    &-\frac{\partial^2}{\partial z^2}E_\mathrm{signal}+\frac{1}{c^2}\frac{\partial^2}{\partial t^2} E_\mathrm{signal} =-\mu_0 N (d_{ee}-d_{gg})\frac{\partial^2}{\partial t^2}\rho_{ee}\\
    &-2\mu_0 N \Re
    \left\{
        d_{ge}
        \left[
            \bessel_1(\kappa)\frac{\partial^2}{\partial t^2}r_{eg} +2\bessel'_1(\kappa)\frac{\partial}{\partial t}\kappa\frac{\partial}{\partial t}r_{eg}\right.\right.
            \left.\left.+
            \left(
                \bessel''_1(\kappa)\left(\frac{\partial}{\partial t}\kappa\right)^2+\bessel'_1(\kappa)\frac{\partial^2}{\partial t^2}\kappa
            \right)
            r_{eg} 
        \right] 
    \right\}.
\end{split}
\label{eq:final}
\end{equation}
Here, $\bessel'_1(\kappa)$ and $\bessel''_1(\kappa)$ are respectively the first and second derivatives of the Bessel function over the argument $\kappa$. These terms represent the influence of the temporal modulations of the drive envelope. For a continuous-wave drive, all time derivatives of $\kappa$ disappear, and the equation is significantly simplified. In general, $\bessel_1'(\kappa)=\frac{1}{2}[\bessel_0(\kappa)-\bessel_2(\kappa)]$ and $\bessel''_1(\kappa)=\frac{1}{4}[\bessel_3(\kappa)-3\bessel_1(\kappa)]$. Because $\kappa$ is typically small, the part proportional to $\bessel_0(\kappa)$ is the dominant contribution. Then, the correction arising from the time dependence of the drive envelope can be rewritten as $\bessel_0(\kappa)\frac{\partial}{\partial t}\kappa \frac{\partial}{\partial t}r_{eg}$. 

From Eq.~(\ref{eq:final}), it follows that both the diagonal and the off-diagonal parts of the density matrix contribute to the generation of low-frequency radiation. Assuming comparable values of permanent and transition dipole moments, we see that the oscillation of the population has a major impact on the signal buildup, in particular in the case of the slowly varying drive envelope, i.e., negligible $\frac{\partial}{\partial t}\kappa$. This observation is one of the key points of this work: the physical origin of the generated signal is the permanent dipole moment associated with the eigenstates of the system and oscillating at the Rabi frequency. This is the carrier frequency of the output signal. The Rabi frequency is determined by the drive amplitude $\mathcal{E}$, which provides a knob for spectral tuning of the signal.

In Eq.(\ref{eq:final}) we assume co-linear orientation of molecules. To include a distribution of the orientation directions of the molecules, one would need to replace the density $N$ with a distribution function $N(\theta,\phi)$, and integrate over the orientations $(\theta,\phi)$. In practice, this would suppress the coherence of the generated signal due to the distribution of coupling strengths between the differently oriented molecules and the drive.

Note that in many other works the field propagation equation in the Bloch-Maxwell set is a first-order differential equation \cite{scully,paspalakis2000,slowik2012}. That simplified form is obtained under the slowly varying envelope approximation in which the signal envelope is assumed to vary both in time and space much more slowly than its inverse carrier frequency and wave vector, respectively. Here, this approximation may not be justified, because the carrier frequency $\Omega_R$ is of the same order as the inverse timescales of the system's dynamics, and therefore we chose to retain the more complicated second-order propagation equation.

\section*{Results \& Discussion}
In this section we first perform calculations for parameters that do not describe any specific molecule, but rather represent orders of magnitude characterizing standard molecular or atomic ensembles, in order to present the typical output provided by the model. To explore the possibilities and limitations, in some investigations we even neglected decoherence or consider very large concentrations. Next, we investigate the performance of a LiH molecular ensemble as a low-frequency signal source.

\subsection*{Model parameters}

\begin{figure}[!b]
\centering
\includegraphics[scale=0.34]{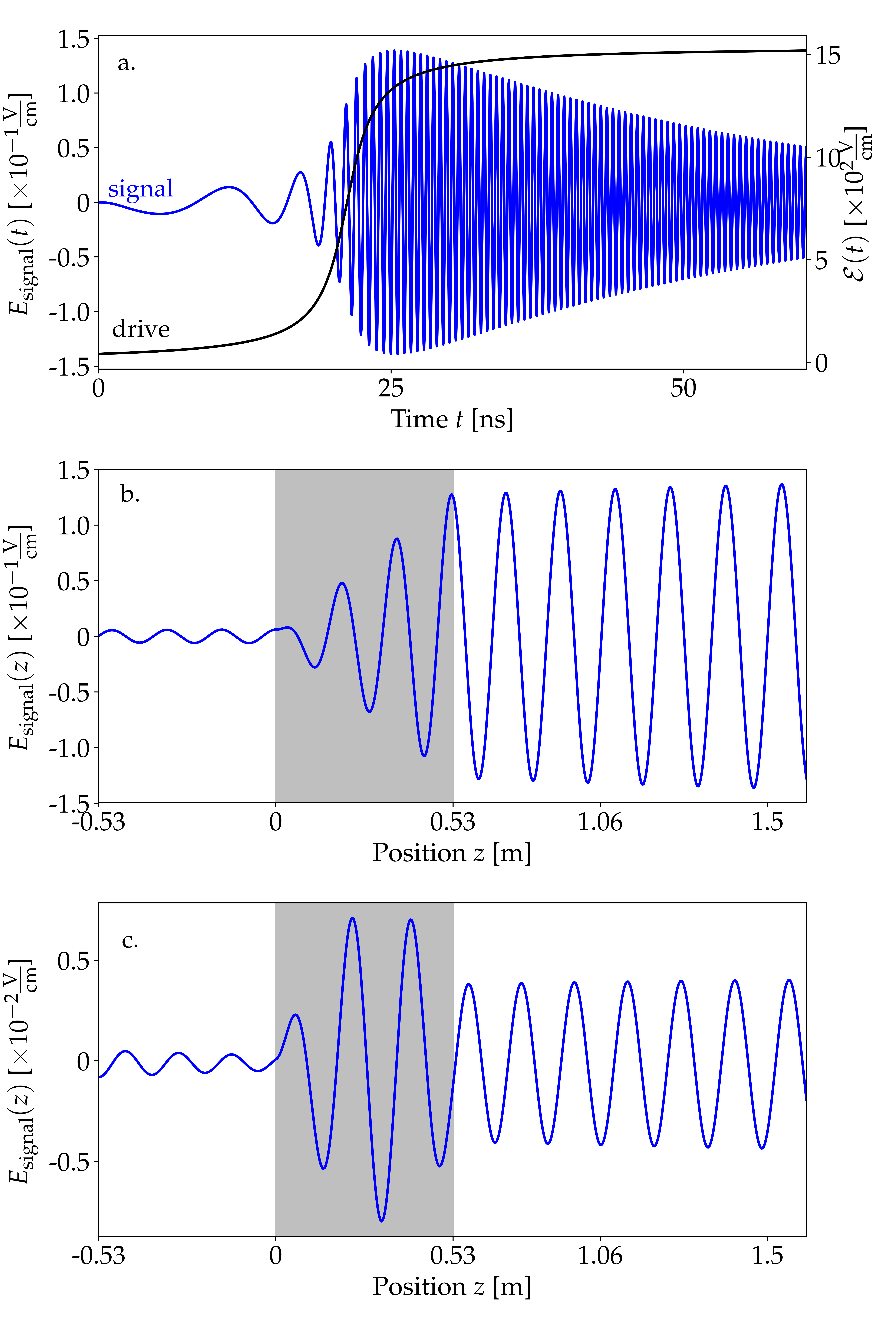}
\caption{\textbf{a}. Envelope of the driving field $\mathcal{E}$ (black) and generated signal $E_\mathrm{signal}$ (blue) as functions of time at the end of the sample ($z=L$) for $\gamma_\mathrm{se}=3.4$~MHz, $\gamma_\mathrm{coll} =65$~kHz, and $\delta =0$.  \textbf{b}. The signal $E_\mathrm{signal}$ in the spatial domain at $t=30$~ns. The gray rectangle represents the sample with the active medium (between 0 and 0.53~m). \textbf{c}. Spatial behavior of the generated radiation for the same $\gamma_\mathrm{se}$ and $\gamma_\mathrm{coll}$ coefficients but in the presence of detuning $\delta=460$~MHz.}
\label{fig:signal}
\end{figure}

We apply the theory to a model sample of a gaseous molecular medium. 
We focus on an electronic transition at $\omega_0=660$~THz, with dipole moments set to 1 atomic unit (\mbox{$d_{ee}=d_{eg}=8.5\times10^{-30}$~Cm}). The spontaneous emission rate $\gamma_\mathrm{se}=3.4$~MHz is calculated according to the Weisskopf--Wigner theorem \cite{scully}, and the collisional decoherence lifetime $\gamma_\mathrm{coll}^{-1}=65$~kHz has been chosen in accordance with experiments on diluted atomic vapors \cite{dziczek2009}. The sample of length $L=53$~cm is illuminated with a driving beam with the envelope
\begin{equation}
    \mathcal{E}(t,z)=A\frac{1}{\pi}\left(\arctan\big(-\alpha(z-z_0-ct)\big)+\frac{\pi}{2}\right),
    \label{impulse}
\end{equation}
where the $\mathrm{arctan}$ function is chosen to model a smooth ramp-up of the beam. The parameter $\alpha=0.019$/cm is a scaling factor controlling the slope and $z_0=-5.3$~m. The amplitude \mbox{$A=1550$~V/cm} corresponds to continuous-wave laser power of $32$~W for a beam area of 1~mm$^2$. The carrier frequency of the drive will be chosen around the medium resonance. 

\begin{figure}[!b]
\centering
\includegraphics[scale=0.34]{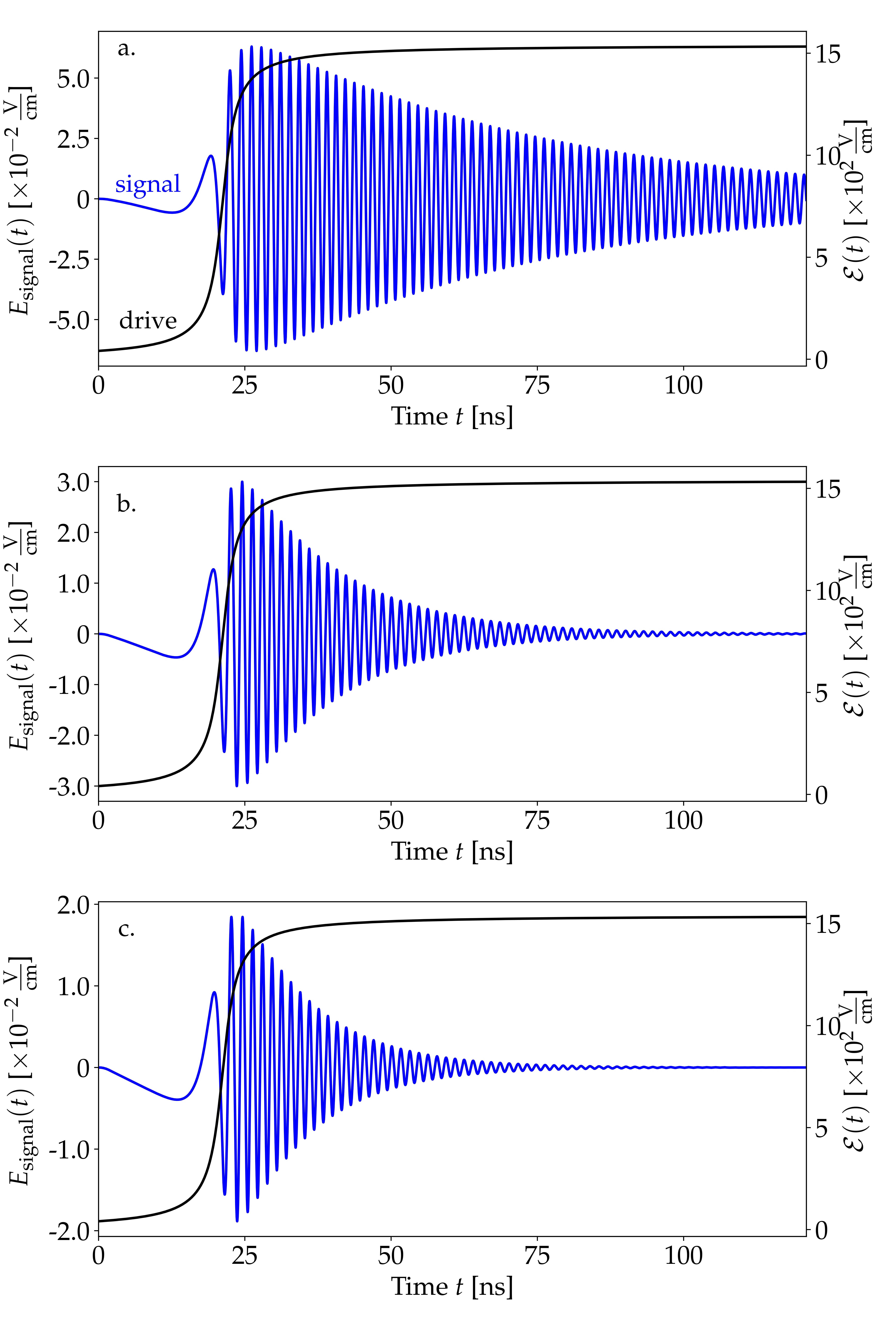}
\caption{Shapes of the signal field (blue) for enhanced values of relaxation parameters. The envelope of the driving field is shown in black. \textbf{a}. Only collisional relaxation present: $\gamma_\mathrm{coll}=6.6$~MHz, $\gamma_\mathrm{se}=0$. \textbf{b}. Only spontaneous emission involved: $\gamma_\mathrm{se}=6.6$~MHz, $\gamma_\mathrm{coll}=0$. \textbf{c}. Full picture: $\gamma_\mathrm{se}=6.6$~MHz, $\gamma_\mathrm{coll}=6.6$~MHz.}
\label{fig:gamma}
\end{figure}

We solve the Bloch--Maxwell equations (\ref{eq:bloch},\ref{eq:final}) 
with a self-developed Python code, which we made available on a public repository\cite{replink}.  The Bloch and the Maxwell equations are alternately iterated in time, so that their coupling is fully accounted for, allowing one in particular to observe the effects of back-action of the signal on a dense medium, as it is discussed later. First, we discuss the case of a relatively low concentration $N=6.7\times10^{12}$ molecules/cm$^3$. The time dependence of the generated low-frequency signal at the position fixed at the end of the sample $z=L$ is shown with the blue line in Fig.~\ref{fig:signal}a. As the drive enters the medium, the signal builds up. Note that the signal frequency $\Omega_R(z=L,t)$ varied in time proportionally to the drive envelope $\mathcal{E}(z=L,t)$ and reached the stable value of 1.97~GHz after approximately 30~ns. The amplitude of the signal is significantly weaker than the drive, but well beyond the detection threshold in the microwave domain \cite{inomata2016,besse2018}. Naturally, the weak amplitude of the signal is a result of the small medium density we chose and could be improved in denser samples. Note that for very large densities at which the mean spatial separation between the molecules would be below the transition wavelength, dipole--dipole interactions of molecules may become relevant, but are not included in our model. They could be taken into account, e.g., through local-field corrections \cite{bowden1993}.

Fig.~\ref{fig:signal}b shows the generated signal at a fixed time \mbox{$E(z,t=30\,\mathrm{ns})$.} The gray area corresponds to $z\in[0,L]$ and represents the active medium, where the amplitude of the signal grows steadily. We find the dominant component of the signal to propagate toward the positive-$z$ direction, in accordance with the propagation direction of the drive. This indicates the coherent character of the generated signal. Outside the sample the signal amplitude retains its value. 

Fig.~\ref{fig:signal}c shows the same dependence as Fig.~\ref{fig:signal}b, but for a drive detuned by $\delta=460$~MHz. Naturally, the generated signal field amplitude is decreased in this case and its frequency shifted, as we will discuss below. In addition, the detuning induces a beating of the signal field inside the sample, now modulated with a sinusoidal envelope of the spatial period corresponding to the detuning. 
In consequence, the amplitude of the output signal outside the sample is suppressed depending on the phase of the envelope of the signal at the end of the sample. 

Due to decoherence, the signal decays at timescales determined mostly by $\gamma_\mathrm{se}^{-1}$, while $\gamma_\mathrm{coll}$ is a relatively less important correction (see Fig. \ref{fig:gamma}). The reason is that the relaxation $\gamma_\mathrm{coll}$ does not directly affect the population (i.e., it does not appear in Eq.~\ref{eq:blocha}), which is the main source of the low-frequency signal. The impact of relaxation is through a modification of the coherence term $r_{eg}$.
The decay of the signal is a direct consequence of the decay of population oscillations. 
This means that decoherence channels, in particular a strong spontaneous emission, might prevent the possibility of low-frequency signal generation in a continuous manner. Instead, the signal could be generated pulse-wise, with drive impulses as discussed at the end of this subsection.

\begin{figure}[!b]
\centering
\includegraphics[scale=0.34]{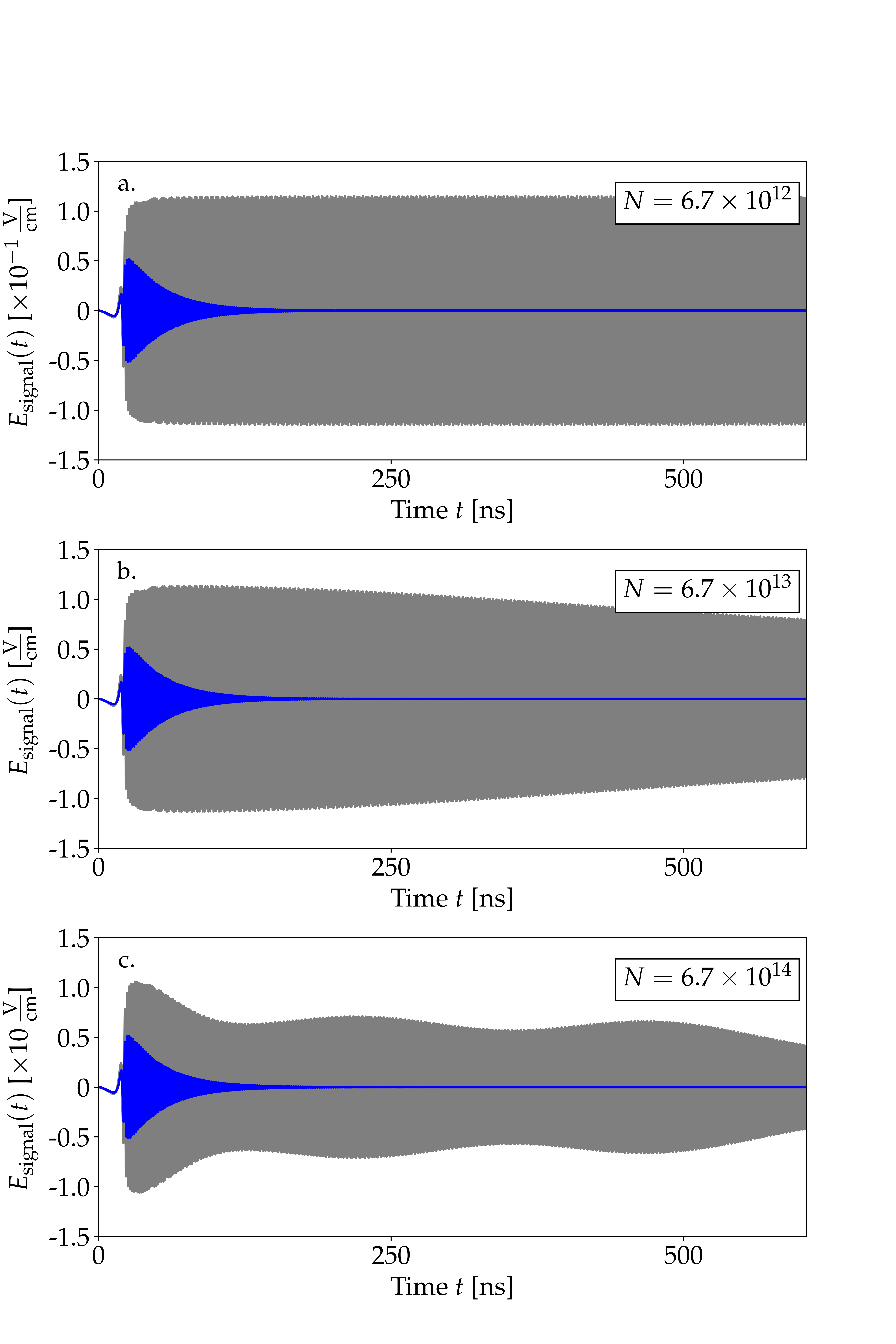}
\caption{Generated signal for different concentrations $N$ in the medium with (blue) and without (gray) relaxations. The driving field's amplitude was set to $A=515$~V/cm. \textbf{a}. Concentration $N=6.7\times10^{12}$ molecules/cm$^3$. \textbf{b}. $N=6.7\times10^{13}$ molecules/cm$^3$. \textbf{c}. $N=6.7\times10^{14}$ molecules/cm$^3$. }
\label{fig:backaction}
\end{figure}

\begin{figure}[!t]
\centering
\includegraphics[scale=0.4]{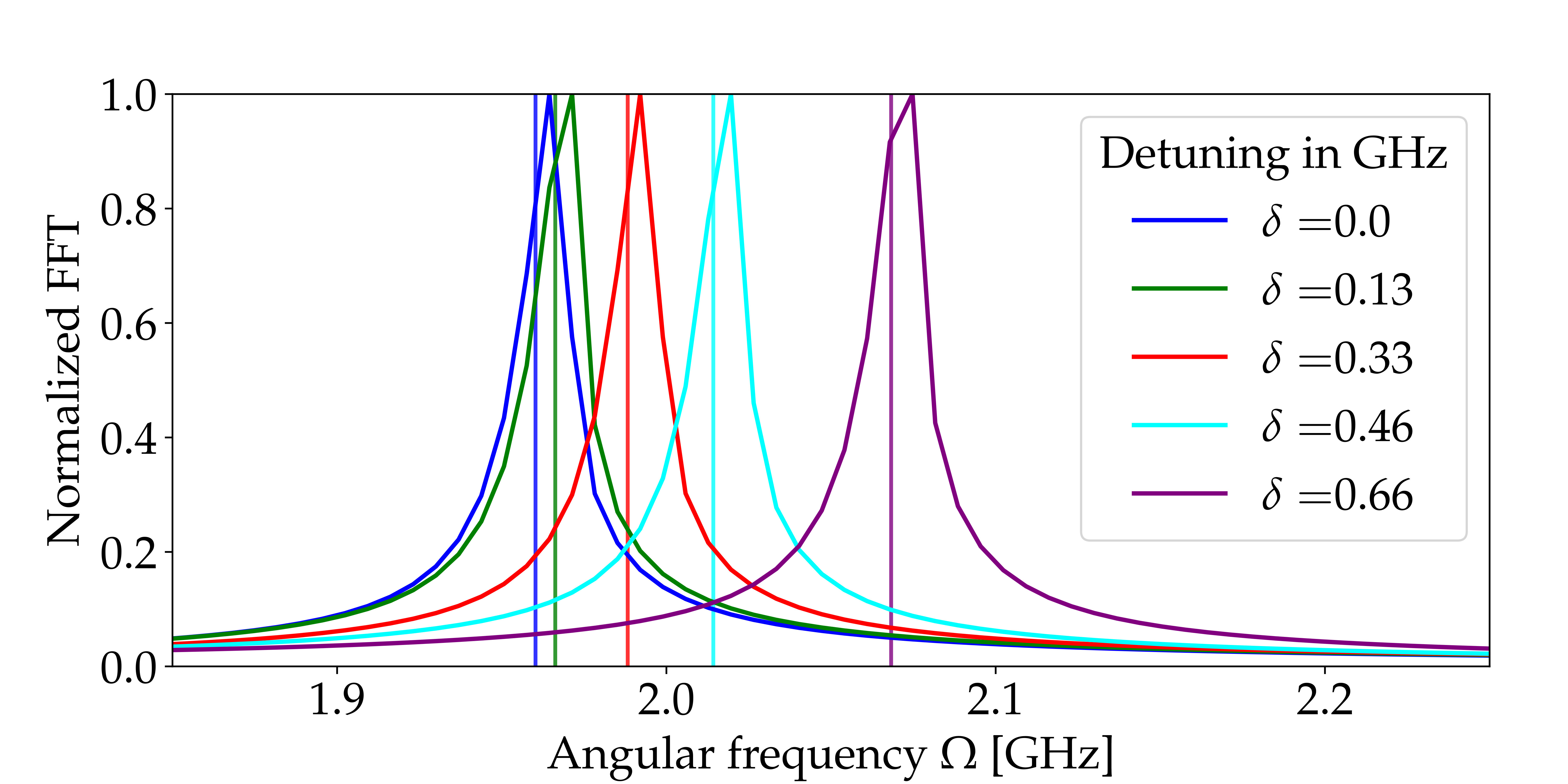}
\caption{Fast Fourier transforms (FFTs) of the generated signals reveal frequencies of the signals for the resonant case (blue line) and for several values of detuning (other lines), all for $\Omega_R=1.97$~GHz. The vertical lines indicate frequencies calculated according to Eq.~(\ref{eq:Rabi_freq_theoretical}).}
\label{fig:fourier}
\end{figure}

A close analysis of Eq.~(\ref{eq:blochb}) reveals that an effective detuning $\delta_\mathrm{eff}(z,t)=\delta-\frac{\partial}{\partial t}\kappa-\frac{E_\mathrm{signal}}{\hslash} (d_{ee}-d_{gg})$ can be induced by additional effects: a time modulation of the drive and a back-action from the generated signal field. While the former is weak for the modulation timescales considered in our examples, we analyze the impact of the signal back-action in samples with increased concentration $N$ in Fig.~\ref{fig:backaction}. The calculations were performed for a weaker drive ($A=514$~V/cm), longer time period, and the same decoherence parameters for better visualisation. In the case of small concentrations, the back-action is not observed at the investigated time scales (Fig.~\ref{fig:backaction}a). For larger medium concentrations, we find a clear indication of the back-action from the signal only if we do not include the spontaneous emission (gray curves in Figs.~\ref{fig:backaction}b,c): the modulations of the signal amplitude originate from time- and space-dependent effective frequency fluctuations ($\delta_\mathrm{eff}$) of the atomic transition induced by the signal. However, this effect is blurred in the presence of spontaneous emission (blue curves in Figs.~\ref{fig:backaction}b,c). The back-action of the signal also appears in Eqs.~(\ref{eq:bloch}) next to the Rabi frequency $\Omega_R$, modifying it to the effective value \mbox{$\Omega_\mathrm{eff}=\Omega_R+\frac{E_\mathrm{signal}}{\hslash}\bessel_1(\kappa)d_{eg}$}. Yet, this impact is considerably smaller due to the proportionality of the correction applied to the Bessel function $\bessel_1(\kappa)\ll 1$.

\begin{figure}[!t]
\centering
\includegraphics[scale=0.35]{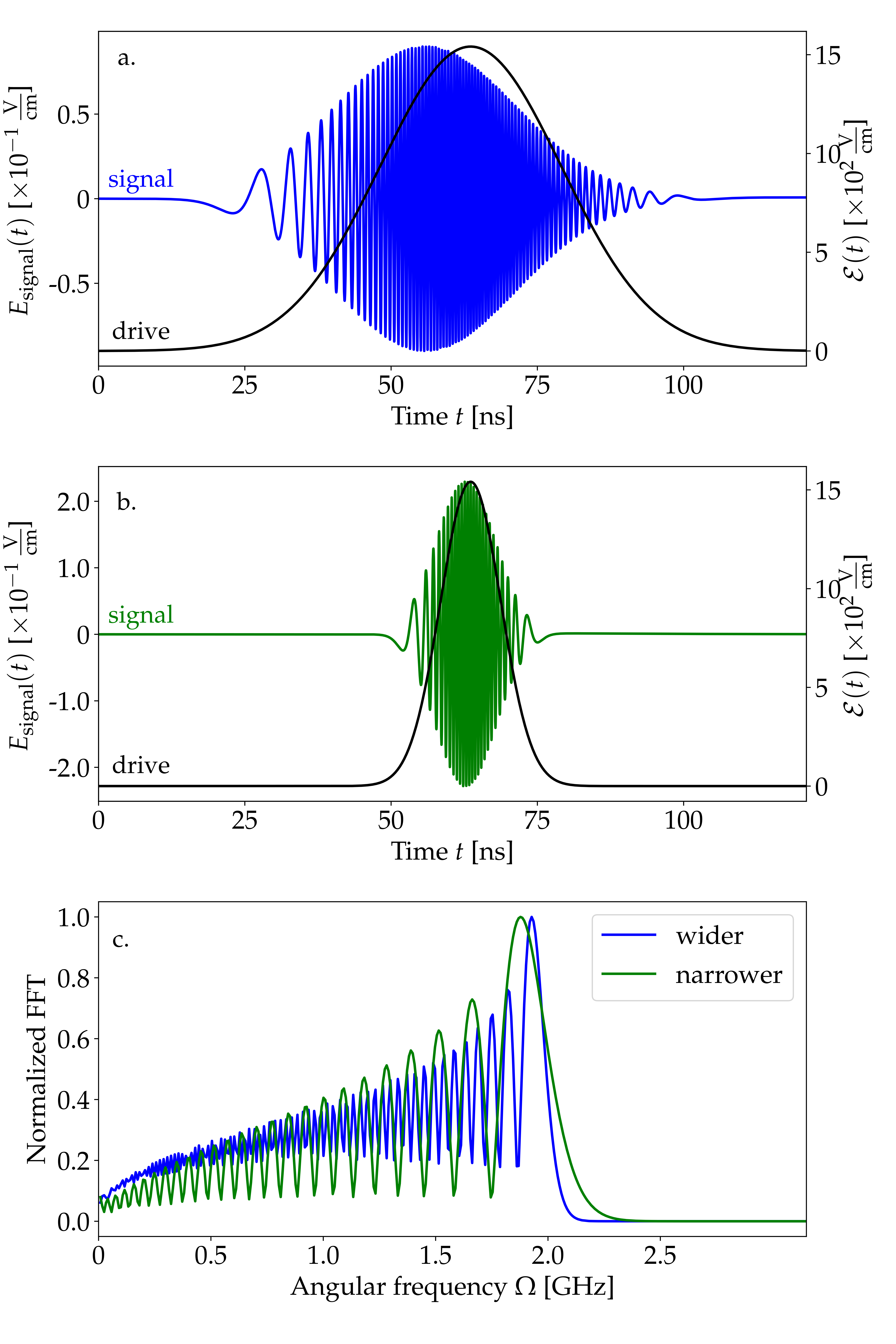}
\caption{Results of the simulation for the impulse driving field with amplitude $A=515$~V/cm. Blue and green lines represent the generated signal while black ones are envelopes of the drive in the form of Gaussians with respective FWHMs of \textbf{a}. 36~ns and \textbf{b}. 12~ns. \textbf{c}. Fast Fourier transform performed on signal impulses.}
\label{fig:gaussy}
\end{figure}

To quantify the impact of the drive detuning on the signal frequency, we evaluate Fourier transforms of the stationary parts of the generated signals (the integration of the signal field $E_\mathrm{signal}(z=L,t)$ was performed over times between 72~ns and 121~ns). Their normalized values are shown in Fig.~\ref{fig:fourier}. We compare the results obtained numerically for different values of detuning up to $\delta=657$~MHz, with the theoretical prediction of the oscillation frequency of medium populations \cite{Loudon} \begin{equation}\label{eq:Rabi_freq_theoretical}
 \Omega=2\sqrt{\Omega^2_R+\delta^2/4}.
\end{equation}
The factor of 2 appears because $\Omega_R$ corresponds to the frequency of oscillations of probability amplitudes, while the population probabilities are modulated at twice the pace. The formula above is valid for the case of vanishing relaxation rates, while in the studied case $\gamma_\mathrm{se,coll}$ are approximately two orders of magnitude smaller than $\delta$. Therefore, the main consequence of relaxation is a spectral broadening of the signal rather than a frequency shift. 
In the above expression, we did not take into consideration the effective parameters $\Omega_\mathrm{eff}$, $\delta_\mathrm{eff}$ because their influence at the relevant timescale is negligible with respect to the impact of spontaneous emission. The theoretical results are in very good agreement with the numerical ones, with offsets at the third significant digit, i.e., exactly at the level at which we expect corrections from the relaxations. 

The derived Bloch--Maxwell equations allow us to study the generation of low-frequency radiation under more complex illumination schemes than a smooth step function. In Fig.~\ref{fig:gaussy}, we present the temporal shape of signals generated under illumination with Gaussian impulses defined as
\begin{equation}
    \mathcal{E}(z,t)=A\euler^{-\alpha(z-z_0-ct)^2},
\end{equation}
where $\alpha=a^2\times2.3\times10^{-6}$/cm$^2$ hence, the full temporal width at half-maximum FWHM($a$)=$a\times12$~ns. 
The signals have Gaussian envelopes. The frequency chirp results from the drive modulation. The shift of the maximum of the signal with respect to the peak of the drive is due to decoherence and disappears in the absence of spontaneous emission and relaxation.
Fig.~\ref{fig:gaussy}c presents fast Fourier transforms (FFTs) of both signals. The main frequencies are $\Omega =1.93$~GHz and 1.88~GHz for wider and narrower impulses, respectively. We observe that for increasing temporal FWHMs of the drive, the main frequency of the signal corresponds to the value of $1.97~$GHz, which is the continuous-wave limit.

\begin{figure}[!t]
\centering
\includegraphics[scale=0.34]{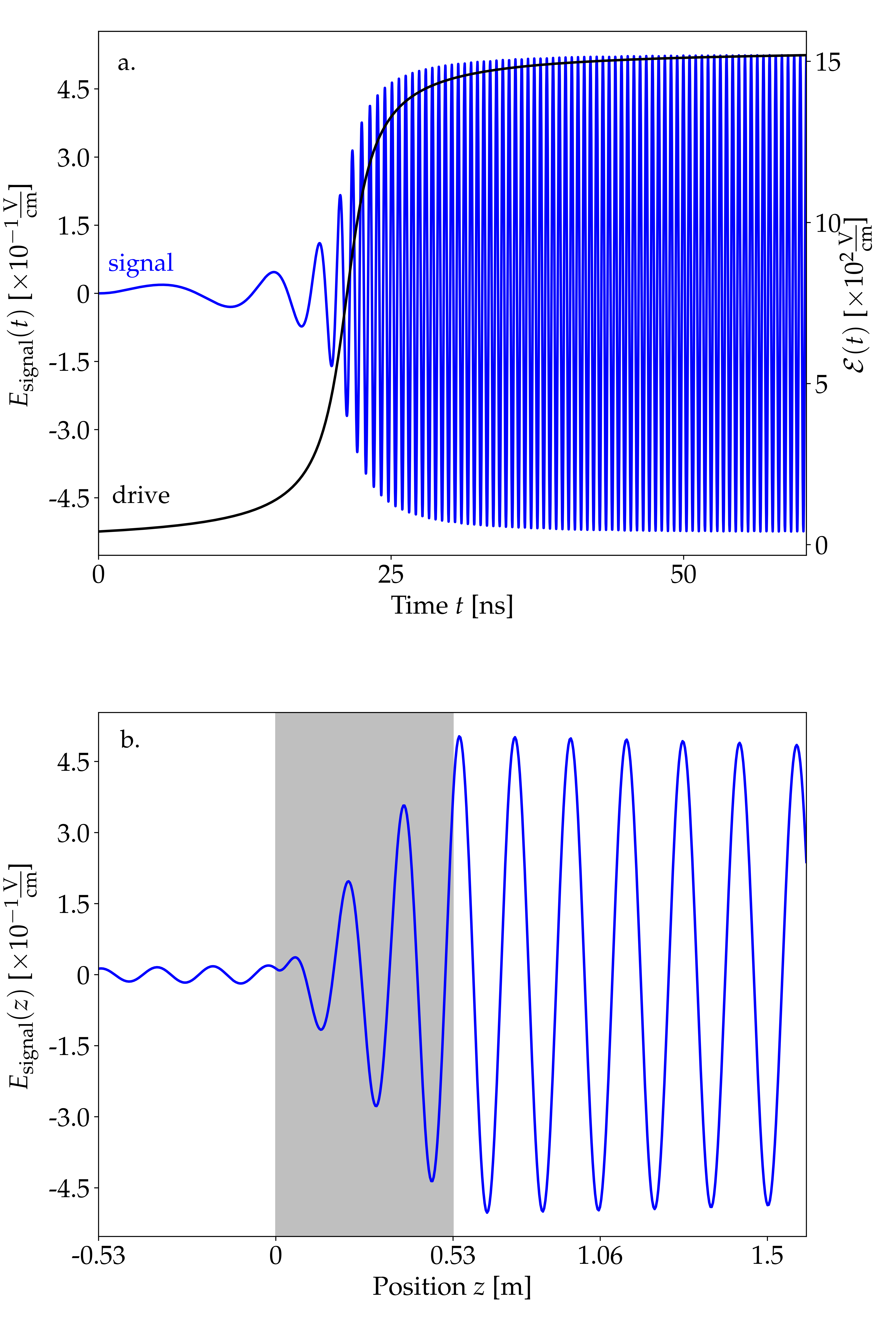}
\caption{\textbf{a}. Envelope of the driving field $\mathcal{E}$ (black) and generated signal $E_\mathrm{signal}$ (blue) as functions of time at the end of the sample ($z=L$) for $\gamma_\mathrm{se}=0.75$~Hz, $\gamma_\mathrm{coll} =65$~kHz, and $\delta =0$. \textbf{b}. The signal $E_\mathrm{signal}$ in the spatial domain at $t=30$~ns. The gray rectangle represents the sample with the active medium (between 0 and 0.53~m).}
\label{fig:sup_LiHoutput}
\end{figure}

Note that a pulsed drive brings into perspective a considerable increase in the driving field power, and therefore the frequency of the signal. In particular, signals in the terahertz domain could be achieved for the drive peak powers from 3.5~MW, assuming the other system parameters as selected above. 

\subsection*{LiH molecule}

In this section we discuss an example of a polar molecule of lithium hydride LiH that well suits the purpose of this project, i.e., it has a large permanent electric dipole moment that can be relatively easily oriented in the laboratory frame with an external electric field. We consider two selected rotational states $\ket{NM}=\ket{00}$ and $\ket{N'M'}=\ket{10}$ of a LiH molecule in its ground vibrational and electronic state $X^1\Sigma^+$. $N$ and $M$ denote the rotational angular momentum of the molecule and its projection on the quantization axis determined by the external DC field. The details of the calculations of the dipole moments (in the laboratory frame) and energies are as a function of applied DC electric field, $E_{DC}$, are provided in the Supplementary Information. At the experimentally achievable electric field  $E_{DC}=150$~kV/cm the energy gap between the levels is $\Delta E=0.642$~THz, the difference between the electric dipole moments in the ground and excited levels reads $d_{ee}-d_{gg}=4.05$~D, and the transition electric dipole moment $d_{eg}=2.51$~D. The concentration and the collisional decoherence rate of the molecules remain the same as before. We deem the concentration realistic with the method described in Ref.~\citeonline{dagdigian1979lih}. The results of our calculations are depicted in Fig.~\ref{fig:sup_LiHoutput}. We find a relatively high amplitude of the generated microwave radiation (the signal beam) of approximately $~0.45$~V/cm. The low energy gap between the levels yields low spontaneous emission coefficient and as a result, a relatively long coherence time of the signal (the decay is not visible at time scales shown in the figure).

\section*{Conclusions}
We have shown how coherent radiation at the Rabi frequency is generated and propagates through a one-dimensional medium consisting of two-level systems with broken inversion symmetry. The underlying mechanism is associated with the Rabi oscillations of the population between the eigenstates induced as the resonant drive illuminates the medium, and of the corresponding permanent dipole moments.

To quantify the effect, we derived Bloch--Maxwell equations governing the dynamics of the system. The equations were solved numerically for sets of parameters representing standard molecular media. The results confirm that the generation of low-frequency radiation is mostly caused by oscillations of the population in the medium, while other contributions can be considered as corrections. The output signal amplitude and frequency can be tuned with drive intensity modulation, allowing for an all-optical control of the signal properties. In our examples, the signal frequency belongs to the microwave regime, while the amplitude is small but detectable. A pulsed illumination scheme allows one to increase both the intensity and frequency of the signal, potentially leading to tunable sources of coherent terahertz radiation.

\bibliography{bibliography}



\section*{Acknowledgements}
The authors thank D. Ziemkiewicz for support on numerical aspects of this work. K.S and P.G. acknowledge the National Science Centre, Poland, grant number 2018/31/D/ST3/01487. P.W. contribution was supported by the National Science Centre, Poland, through Project No. 2019/35/B/ST2/01118.

\section*{Author contributions}
P.G. and K.S. equally contributed to idea conception, theory development and results analysis. P.G. implemented the code and performed all simulations. P.W. provided support on molecular physics. All authors contributed to the preparation of the manuscript.

\section*{Competing interests}
Te authors declare no competing interests.

\section*{Additional information}
\noindent\textbf{Supplementary information} is available for this paper at \href{https://doi.org/10.1038/s41598-020-74569-w}{https://doi.org/10.1038/s41598-020-74569-w}.

\noindent\textbf{Correspondence} and requests for materials should be addressed to P.G.

\noindent \textbf{The numerical code} is available at a public repository (Ref. \citeonline{replink}).


\end{document}


\maketitle
\thispagestyle{empty}

\section{Relation between broken inversion symmetry and dipole moment}    
A diagonal element of the dipole moment operator with $i=e,g$ is given by
\begin{equation}
    \mathbf{d}_{ii}=\bra{i}\hat{\mathbf{d}}\ket{i}=\int d \mathbf{r} \;\sum_\alpha q_\alpha\mathbf{r}_\alpha \bra{i}\ket{\mathbf{r}}\bra{\mathbf{r}}\ket{i},
\label{eq:sup_dii}    
\end{equation}
where $\mathbf{r}=\{\mathbf{r}_\alpha\}_{\alpha=1,\dots,N}$ represents the set of positions $\mathbf{r}_\alpha$ of all $N$ charges $q_\alpha$ contributing to the dipole moment of the system. 
For simplicity, in this proof we assume a single charge $q$ at position $\mathbf{r}_1$. The proof for multiple charges is a straightforward generalization. 

\begin{figure}[ht]
\centering
\includegraphics[scale=0.5]{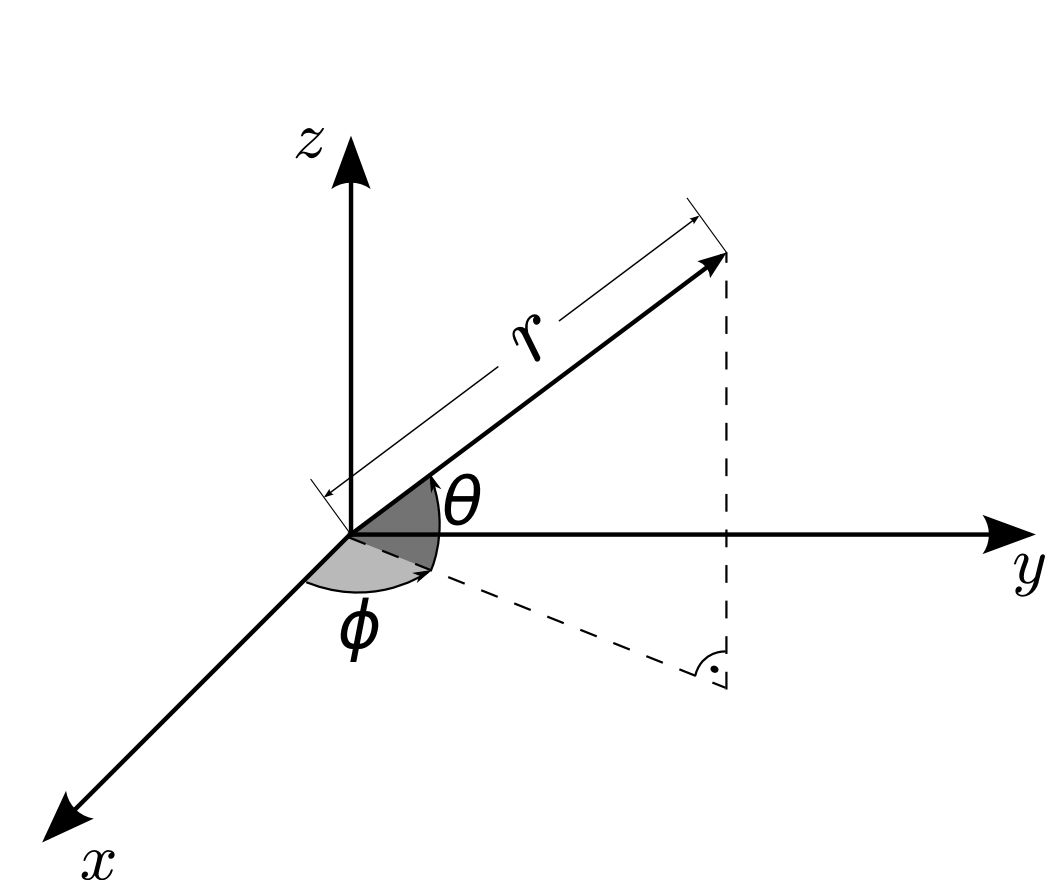}
\caption{Spherical coordinate system used in calculations according to (\ref{eq:sup_coord}). The unusual choice of directed angles $\theta$ and $\phi$ (notice the arrows in the picture) was intentional to set symmetric limits of integration in the equations.}
\label{fig:sup_Coords}
\end{figure}

To evaluate this quantity, we express $\mathbf{r}_1$ in spherical coordinates defined as follows (Fig. \ref{fig:sup_Coords}):
\begin{equation}
\begin{split}
    &x=r\cos\phi\cos\theta,\\
    &y=r\sin\phi\cos\theta,\\
    &z=r\sin\theta.
\end{split}    
\label{eq:sup_coord}
\end{equation}
to obtain
\begin{equation}
\begin{split}
    \mathbf{d}_{ii}&=q\int_{0}^{\infty} dr_1 \int_{-\pi}^{\pi} d\phi_1 \int_{-\pi/2}^{\pi/2} d\theta_1 \;r_1^2\cos\theta_1 \;\mathbf{r}_1\; \abs{\bra{\mathbf{r}_1}\ket{i}}^2\\
    &=q\int_{0}^{\infty} dr_1 \int_{-\pi}^{\pi} d\phi_1 \int_{0}^{\pi/2} d\theta_1 \;r_1^2\cos\theta_1 \;\mathbf{r}_1 \;\abs{\bra{\mathbf{r}_1}\ket{i}}^2 + q\int_{0}^{\infty} dr_1 \int_{-\pi}^{\pi} d\phi_1 \int_{-\pi/2}^{0} d\theta_1 \;r_1^2\cos\theta_1 \;\mathbf{r}_1 \;\abs{\bra{\mathbf{r}_1}\ket{i}}^2
\end{split}    
\label{eq:sup_d}
\end{equation}
By substituting angles $\phi\rightarrow -\phi$, $\theta\rightarrow -\theta$ in the second term and making use of the relation $\cos(-\theta)=\cos(\theta)$, we obtain
\begin{equation}
    \mathbf{d}_{ii}=q\int_{0}^{\infty} dr_1 \int_{-\pi}^{\pi} d\phi_1 \int_{0}^{\pi/2} d\theta_1 \;r_1^2\cos\theta_1\; \mathbf{r}_1
    \bigg( \abs{\bra{\mathbf{r}_1}\ket{i}}^2 - \abs{\bra{-\mathbf{r}_1}\ket{i}}^2\bigg),
\label{eq:sup_ddiag}
\end{equation}
where $-\mathbf{r}_1=(r_1,-\phi_1, -\theta_1)$. 
If the wave function $\bra{\mathbf{r}_1}\ket{i}$ has inversion symmetry, $\abs{\bra{\mathbf{r}_1}\ket{i}} = \abs{\bra{-\mathbf{r}_1}\ket{i}}$ holds, and the bracket at the end disappears. This implies that the diagonal elements $d_{ee}$ and $d_{gg}$ are equal to zero. This means that broken inversion symmetry is required to create a permanent dipole moment.

For the off-diagonal elements, we simply have
\begin{equation}
\begin{split}
    \mathbf{d}_{ij}&=\bra{i}\hat{\mathbf{d}}\ket{j}=
    \int \dd \mathbf{r}\; \mathbf{r} \bra{i}\ket{\mathbf{r}}\bra{\mathbf{r}}q\hat{\mathbf{r}} \ket{\mathbf{r}_1}\bra{\mathbf{r}_1}\ket{j}=\mathbf{d}_{ji}^\star,
\end{split}
\label{eq:sup_doffdiag}
\end{equation}
regardless of the wave function's symmetry. Due to the odd character of the position operator, if both wave functions share the same symmetry, the integral is equal to zero, and the transition is electric-dipole-forbidden.

\section{Bloch--Maxwell equations}
Here, we provide a detailed derivation of the Bloch--Maxwell equations. First, let us insert the form of the field given by Eq.~(1) into the master equation~(6). The set of equations for the density matrix elements has the following form:
\begin{subequations}
\begin{equation}
    \im\hslash\frac{\partial}{\partial t}\rho_{ee}=\left(\mathcal{E} \cos(kz-\omega t)+E_\mathrm{signal} \right) \big(d_{ge} \rho_{eg} - d_{eg} \rho_{ge} \big)
    -2\im\hslash\gamma_\mathrm{se}\rho_{ee},
    \label{eq:sup_blochee} 
\end{equation}
\begin{equation}
    \im\hslash\frac{\partial}{\partial t}\rho_{eg}=\hslash \omega_0 \rho_{eg} -\left(\mathcal{E} \cos(kz-\omega t)+E_\mathrm{signal} \right)
    \big( (d_{ee} - d_{gg})\rho_{eg} - d_{eg}(2\rho_{ee}-1) \big)-\im\hslash(\gamma_\mathrm{se}+\gamma_\mathrm{dec})\rho_{eg}.
    \label{eq:sup_blocheg} 
\end{equation}
The dynamics of the other elements can be found based on the density matrix properties: $\rho_{ee}+\rho_{gg}=1$ and $r_{ge}=r_{eg}^\star$.
\end{subequations}
Next, we make use of the ansatz~(9) and expand the $\kappa$-dependent term into a series of Bessel functions $e^{-\im\kappa\sin x}=\sum_{n=-\infty}^{+\infty}\bessel_n(\kappa)e^{-\im nx}$:
\begin{subequations}
\begin{equation}
    \im\hslash\frac{\partial}{\partial t}\rho_{ee}
    =\left[\frac{1}{2}\mathcal{E}\left(\euler^{2\im(kz-\omega t)}+1\right)+E_\mathrm{signal}\euler^{\im(kz-\omega t)}\right]
    \sum_{n=-\infty}^{\infty}\bessel_n(\kappa)\euler^{-\im n (kz-\omega t)}d_{eg}^\star r_{eg}+\mathrm{c.c.}-2\im\hslash\gamma_\mathrm{se} \rho_{ee},
    \label{eq:sup_blochee_full}
\end{equation}
\begin{equation}
    \begin{split}
    \im\hslash\frac{\partial}{\partial t}r_{eg}
    &=\hslash\left[\delta-\frac{\partial}{\partial t}\kappa-E_\mathrm{signal}(d_{ee}-d_{gg})\right]r_{eg}\\
    &+\left[\frac{1}{2}\mathcal{E}\left(\euler^{2\im (kz-\omega t)}+1\right)+E_\mathrm{signal}\euler^{-\im(kz-\omega t)}\right]d_{eg}
    (2\rho_{ee}-1)\sum_{n=-\infty}^{+\infty}\bessel_n(\kappa)\euler^{\im n (kz-\omega t)}-\im(\gamma_\mathrm{es}+\gamma_\mathrm{dec})r_{eg}.
    \label{eq:sup_blocheg_full} 
    \end{split}
\end{equation}
\end{subequations}
If $\Omega_R\ll\omega$, the oscillatory terms on the right-hand side of the above equations make negligible contributions. We neglect them under the rotating wave approximation, in which out of the infinite sums the only surviving terms correspond to $n=0,2$ if they are multiplied by the drive envelope $\mathcal{E}$, or correspond to $n=1$ if they are multiplied by the signal field. We arrive at the equations~(10) from the article.

To derive formula~(12), we insert Eq.~(1) into the right-hand side of Eq.~(4). Performing all the derivatives, we obtain
\begin{equation}
\begin{split}
    -\frac{\partial^2}{\partial z^2} E+\frac{1}{c^2} \frac{\partial^2}{\partial t^2} E&=-\frac{\partial^2}{\partial z^2}E_\mathrm{signal}+\frac{1}{c^2}\frac{\partial^2}{\partial t^2}E_\mathrm{signal}\\
    &+\frac{1}{2}\left(-\frac{\partial^2}{\partial z^2}+\frac{1}{c^2}\frac{\partial^2 }{\partial t^2}-2\im k \frac{\partial}{\partial z}+2\im\frac{\omega}{c^2}\frac{\partial}{\partial t}+k^2-\frac{\omega^2}{c^2}\right)\mathcal{E}
    \left( e^{\im(kz -\omega t)}-\euler^{-\im(kz -\omega t)}\right).
\end{split}
\label{eq:sup_field}
\end{equation}
Clearly, the slowly varying terms of the expression correspond to the signal, while the terms related to the drive envelope are multiplied by rapidly oscillating factors. 

Let us now investigate the right-hand-side of the wave equation with the second time derivative of polarization. The fastest method is to apply the derivatives to the form of polarization given by Eq.~(11). We introduce the symbol
\begin{equation}
    R=\sum_{n=-\infty}^{\infty}\bessel_n(\kappa)\euler^{-\im(n-1)(kz-\omega t)}.
\label{eq:sup_R}    
\end{equation}
The second time derivative of the polarization is
\begin{equation}
\frac{\partial^2}{\partial t^2}P=N\left[ (d_{ee}-d_{gg})\frac{\partial^2}{\partial t^2}\rho_{ee} +2\frac{\partial^2}{\partial t^2}\Re(d_{eg}^\star r_{eg}R)\right].
\label{eq:sup_pol}
\end{equation}
The second term in the bracket reads
\begin{equation}
\begin{split}
    \frac{\partial^2}{\partial t^2} (d_{eg}^\star r_{eg}R)&=d_{eg}^\star\left(R\frac{\partial^2}{\partial t^2}r_{eg}+2\frac{\partial}{\partial t}r_{eg}\frac{\partial}{\partial t}R+r_{eg}\frac{\partial^2}{\partial t^2}R\right)\\
    &=\sum_{n=-\infty}^{\infty}\Bigg[\Bigg( \frac{\partial^2}{\partial t^2}r_{eg} +2\frac{\partial}{\partial t}r_{eg}\Big(\bessel'_n(\kappa)\frac{\partial}{\partial t}\kappa +\im(n-1)\omega\bessel_n(\kappa)\Big)
    +r_{eg}\Big( \bessel''_n(\kappa)\Big(\frac{\partial}{\partial t}\kappa\Big)^2+ \bessel'_n(\kappa)\frac{\partial^2}{\partial t^2}\kappa\\
    &+2\im(n-1)\omega \bessel'_n(\kappa)\frac{\partial}{\partial t}\kappa +[\im(n-1)\omega]^2\bessel_n(\kappa)\Big)\Bigg)d_{eg}^\star\euler^{-\im(n-1)(kz-\omega t)}\Bigg],
\end{split}
\label{eq:sup_pol_full}
\end{equation}
where we have explicitly inserted the first and second derivatives of $R$ in accordance with Eq.~(\ref{eq:sup_R}). We insert this result back into Eq.~(\ref{eq:sup_pol}).

Now, in analogy with the procedure conducted for the left-hand side (\ref{eq:sup_field}) of the wave equation, we can separate on the right-hand-side terms proportional to different powers of the oscillating factor $\mathrm{exp}[i(kz-\omega t)]$.
In the rotating wave approximation, we neglect powers other than 0 and 1, as only these two are present on the left-hand side given by Eq.~(\ref{eq:sup_field}).
Now, we make the important assumption that any cross-talk between the slowly varying terms and those oscillating at the frequency $\omega$ can be neglected: The polarization terms oscillating at $\omega$ are coupled to the drive, while the slowly varying terms act as the source for the signal. This assumption is valid if $\left|\frac{\partial \mathcal{E}_\mathrm{signal}}{\partial t}\right|\ll\omega\mathcal{E}_\mathrm{signal}$ and similarly for the slowly varying part of the polarization. As a result, we can separate two wave equations, describing respectively the dynamics of the drive and of the signal. However, we assume the drive to be strong enough not to be affected by the coupling to the medium. The only relevant field equation is therefore the one for the drive, given by~(12). On its right-hand side appear the terms corresponding to $n=1$ in Eq.~(\ref{eq:sup_pol_full}). 

\section{Numerical approach}
The solution of the coupled set of Bloch--Maxwell equations~(10, 12) can be found numerically. 

The wave equation has the form
\begin{equation}
    -\frac{\partial^2}{\partial z^2} f(z,t) +\frac{1}{c^2} \frac{\partial^2}{\partial t^2} f(z,t) = s(z,t),
\label{eq:sup_wave_s}
\end{equation}
where $f$ is the function we wish to find, $s$ is a source term, and $c$ is the speed of the envelope in vacuum. Introducing discretization, we end up with a time-space grid of evenly distributed points $(z_j, t_i)$ with respective spatial and temporal steps $\Delta z$ and $\Delta t$. Hence, for the space variable, we have $z_j=z_0+j\Delta z$, where $j \in [0, 1, ..., N_L]$, and $N_L$ is the number of the point at the end of the sample of length $L$, so $N_L\Delta z=L$. Similarly, $t_i=t_0+i\Delta t$ for $i\in [0, 1, ...]$ but with no upper bound. For convenience, we denote values of the functions at grid points $f(z_j, t_i)\equiv f_{i,j}$ and $s(z_j, t_i)\equiv s_{i,j}$. By the solution at time $t_{i+1}$, we understand the set of values $\{f_{i+1,j}\}$ for all $j$, and hence we have to find an expression that depends only on the previously calculated values. This can be done by expressing the second-order derivatives by the three-point (midpoint) formula \cite{1}. Because we solve a second-order partial differential equation, two initial conditions for each space point are required. We introduce values $f_{j,0}$, $s_{j,0}$ and velocities $\partial f_{j,0}/\partial t \equiv g_j$ for the initial time $t_0$. To reach an accuracy on the order of $(\Delta t)^2$, we express $f_{1,j}$ as \cite{2}
\begin{equation}
    f_{1,j}=\frac{1}{2} \eta^2 (f_{0,j-1}+f_{0,j+1})+(1-\eta^2)f_{0,j}+\Delta t g_j+\frac{1}{2}c^2(\Delta t)^2 s_{0,j},
    \quad \text{for } j \neq 0,N_L,
\end{equation}
and we find
\begin{equation}
    f_{i+1,j}=\eta^2(f_{i,j+1}+f_{i,j-1})+2(1-\eta^2)f_{i,j}-f_{i-1,j}+c^2(\Delta t)^2 s_{i,j},
    \quad \text{for } i>0, j \neq 0,N_L.
\label{eq:sup_inside}
\end{equation}
Obviously, the foregoing expressions are not valid at the ends of the sample. There, we apply transparent boundary conditions \cite{3} for the radiation exiting the sample (for $j=0,N_L$):
\begin{subequations}
\begin{equation}
    f_{1,0/N_L}=\eta^2f_{0,1/N_L-1}+(1-\eta^2)f_{0,0/N_L}+(1-\eta)\Delta t g_{0/N_L},
\label{eq:sup_boundarya} 
\end{equation}
\begin{equation}
    f_{i+1,0/N_L}=\frac{2\eta^2f_{i,1/N_L-1}+2(1-\eta^2)f_{i,0/N_L}+(\eta-1)f_{i-1,0/N_L}}{1+\eta},
\label{eq:sup_boundaryb} 
\end{equation}   
\label{eq:sup_boundary}
\end{subequations}

\noindent where we assumed no sources at the ends of the sample. In addition, we have introduced the parameter $\eta=c\Delta t/\Delta z$, the so-called Courant number first described in Ref.~\citeonline{4}, whose value is crucial for numerical stability. In our case, $\eta=1$ corresponds to the so-called "magic step" \cite{5} and leads to a solution that is not affected by the numerical dispersion problem.

The source term $s(z,t)$ corresponds to the solution of the Bloch Eqs.~(10). The latter is a set of first-order differential equations of one variable for each point in space. At each time step, we calculate the solution using the Python build-in method \textit{odeint} from the \textit{scipy.integrate} library. It is a RK4 method, snd so the order of accuracy is  $(\Delta t)^4$. This high level of accuracy is required because the source term is eventually differentiated twice, reducing the accuracy to the order of $(\Delta t)^2$. The differentiation is performed using the three-point (midpoint) method.

Our solver is written in Python 2.7, where two sets of equations (\ref{eq:sup_inside}, \ref{eq:sup_boundary}) were implemented. A comparison between Eqs.~(12) and (\ref{eq:sup_wave_s}) reveals that $f\equiv E_\mathrm{signal}$, and $s$ is the right-hand side of Eq.~(12). The solver is available on a public repository\cite{6}. We draw the readers attention to another existing solver\cite{7}.

\section{Lithium hydride example}
\begin{figure}[h!]
\centering
\includegraphics[scale=0.49]{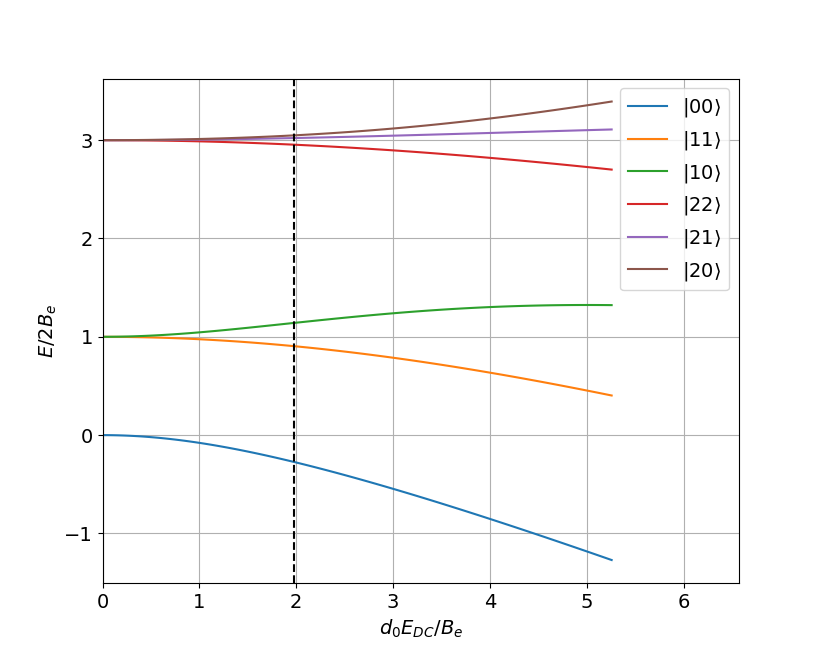}
\caption{Energies of the rotational states in the ground electronic and vibrational state of the LiH molecule, as a function of electric field. The dashed line represents $E_{DC}=150$~kV/cm. Levels represented by the blue and green lines were selected for the calculations in the main text.}
\label{fig:sup_DCenergies}
\end{figure}

\begin{figure}[ht!]
\centering
\includegraphics[scale=0.49]{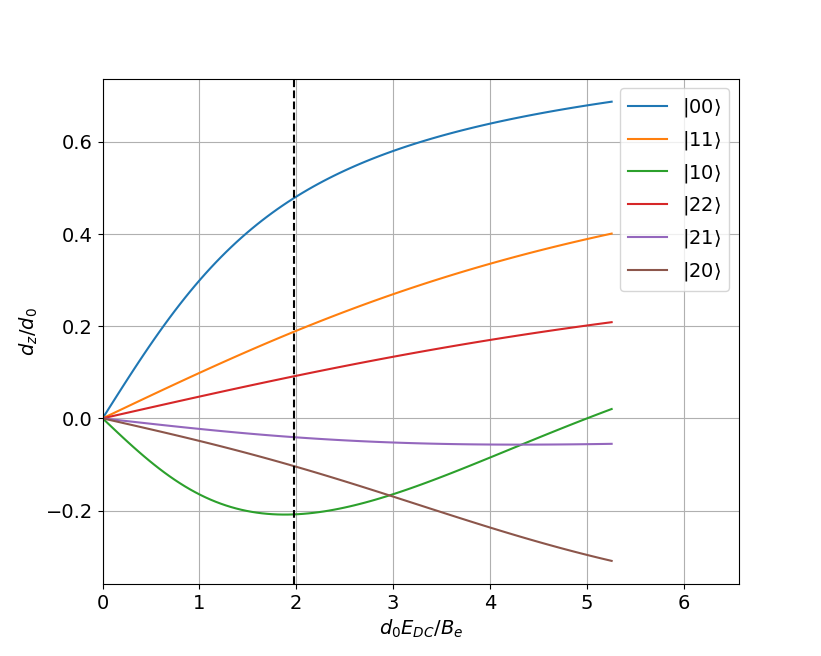}
\caption{Projections of the permanent dipole moments in different rotational states as functions of applied electric field. The dashed line represents the amplitude $E_{DC}=150$~kV/cm. The blue and green curves represent results for the states selected for the calculations in the main text.}
\label{fig:sup_DCdipole}
\end{figure}

To perform calculations based on a real system we have chosen the LiH molecule in ground state $X^1\Sigma^+$. We investigate its rotational states $\ket{NM}$. Since in the ground state the electronic angular momentum is zero, $N$ and $M$ denote the rotational angular momentum of the molecule and its projection. The permanent electric dipole moment in the molecule's frame is $d_0=5.88$~D \cite{8} and the rotational constant $B_e=7.513$~cm$^{-1}$ \cite{9}. In general, the orientation of a molecule in the gaseous ensemble is random and as the result, in the lab frame, the average dipole moment cancels out. To distinguish one of the directions (e.g. z-axis in the lab frame) and orient molecules we may apply an additional, constant electric field $E_{DC}$. This results in the DC Stark effect which leads to mixing of rotational states. The coupling matrix elements between states $\ket{NM}$ and $\ket{N'M'}$ read\cite{10}
\begin{equation}
    V_{NM;N'M'}=-E_{DC}d_0\sqrt{\frac{4\pi}{3}}\bra{NM}\sph_{10}\ket{N'M'},
\end{equation}
where the spherical harmonic $\sph_{10}$ has been used. The $3-j$ Wigner symbols may now be used to evaluate the transition elements. As a result, the following matrix representation of the Hamiltonian is obtained
\begin{equation}
\begin{split}
\ket{00}\qua\ket{10}\qua\ket{20}\qua\ket{30}\qua\ket{11}\qua\ket{21}\qua\ket{31}\qua\ket{22}\qua\ket{32}\qua\ket{33}\quad\\
H_\mathrm{stark}=
\begin{matrix}
\ket{00}\\
\ket{10}\\
\ket{20}\\
\ket{30}\\
\ket{11}\\
\ket{21}\\
\ket{31}\\
\ket{22}\\
\ket{32}\\
\ket{33}
\end{matrix}
\begin{bmatrix}
0& E_{DC}d_1& 0& 0& 0& 0& 0& 0& 0& 0 \\
E_{DC}d_1& 2B_e& E_{DC}d_3& 0& 0& 0& 0& 0& 0& 0 \\
0& E_{DC}d_3& 6B_e& E_{DC}d_6& 0& 0& 0& 0& 0& 0\\
0& 0& E_{DC}d_6& 12B_e& 0& 0& 0& 0& 0& 0\\
0& 0& 0& 0& 2B_e& E_{DC}d_2& 0& 0& 0& 0 \\
0& 0& 0& 0& E_{DC}d_2& 6B_e& E_{DC}d_5& 0& 0& 0 \\
0& 0& 0& 0& 0& E_{DC}d_5& 12B_e& 0& 0& 0 \\
0& 0& 0& 0& 0& 0& 0& 6B_e & E_{DC}d_4& 0\\
0& 0& 0& 0& 0& 0& 0& E_{DC}d_4 & 12B_e& 0\\
0& 0& 0& 0& 0& 0& 0& 0& 0 & 12B_e
\end{bmatrix},
\end{split}
\label{stark}
\end{equation}
where $d_1=-\frac{1}{\sqrt{3}}d_0$, $d_2=-\frac{1}{\sqrt{5}}d_0$, $d_3=-\frac{2}{\sqrt{15}}d_0$, $d_4=-\frac{1}{\sqrt{7}}d_0$, $d_5=-\frac{2\sqrt{2}}{\sqrt{35}}d_0$, $d_6=-\frac{3}{\sqrt{35}}d_0$.

\begin{figure}[ht!]
\centering
\includegraphics[scale=0.49]{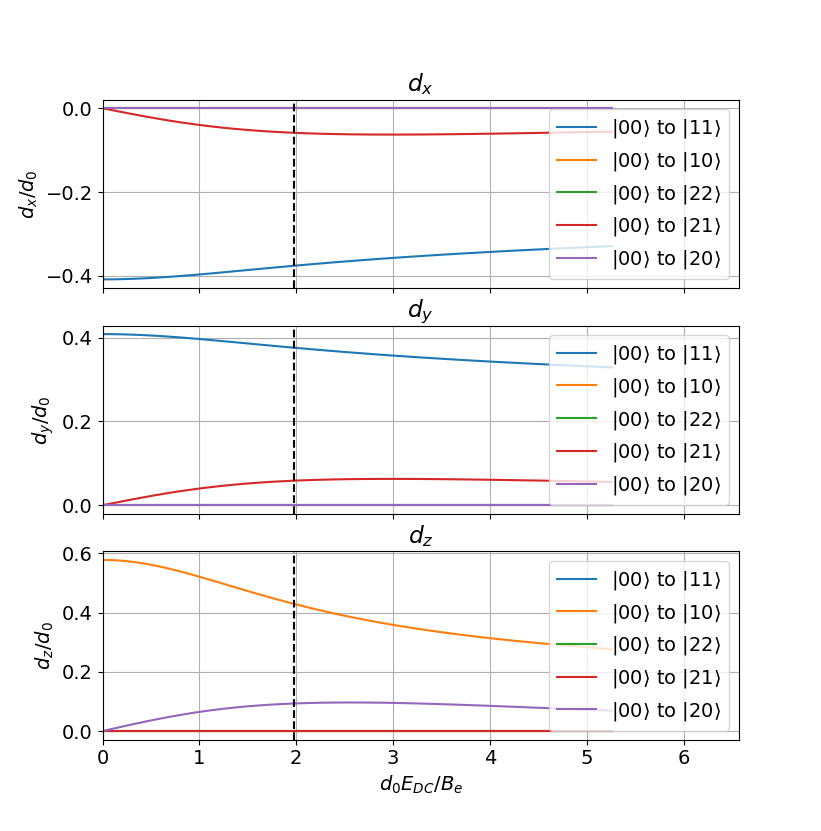}
\caption{Projections on the laboratory frame axes of the transition electric dipole moments for transitions between the ground and selected excited states. The dashed line represents the amplitude $E_{DC}=150$~kV/cm. The orange curve representes the results for the pair of states selected for the calculations in the main text.}
\label{fig:sup_DCtrdipole}
\end{figure}
A diagonalization of the presented Hamiltonian allows us to find new eigenstates and eigenenergies of the system for different values of the DC electric field. The new eigenstates are superpositions of the original states $\ket{NM}$, and for the studied range of fields each eigenstate has one dominant contribution. The eigenenergies are presented in Fig. \ref{fig:sup_DCenergies} for the amplitude $E_{DC}$ in the range between 0 and 400~kV/cm. The label corresponds to the state $\ket{NM}$ whose contribution to the eigenstate dominates. As can be seen, the energy gap between the levels of interest grows with the electric field in the considered range. 

The new eigenstates are characterized with a nonzero permanent electric dipole moment in the lab frame, oriented along the $z$-axis (the direction of the $E_{DC}$) as presented in Fig. \ref{fig:sup_DCdipole}. The black dashed line indicates the field amplitude used in the main text. 
Additionally, transition dipole moments between pairs of states can be induced. In Fig. \ref{fig:sup_DCtrdipole} we present all components of the transition dipole moments between the ground state to a set of possible excited states. As expected, for $M=M'=0$ the transition dipole moments are parallel to the $z$-axis while transitions between levels with $M'\neq 0$ correspond to dipoles in oriented in the $xy$ plane. These dipole moments decrease with the growing field due to the increasing contribution of states different than the dominant one.
